\documentclass[fleqn,usenatbib]{mnras}

\usepackage{newtxtext,newtxmath}
\usepackage[T1]{fontenc}
\DeclareRobustCommand{\VAN}[3]{#2}
\let\VANthebibliography\thebibliography
\def\thebibliography{\DeclareRobustCommand{\VAN}[3]{##3}\VANthebibliography}

\usepackage{graphicx}
\usepackage{amsmath}

\usepackage{lineno}

\title[Shock residual energy]{Residual energy of magnetohydrodynamic shocks}

\author[S. W. Good et al.]{S. W. Good,$^{1}$\thanks{E-mail: simon.good@helsinki.fi}
K. J. Palmunen,$^{1}$
C. H. K. Chen,$^{2}$
E. K. J. Kilpua,$^{1}$
T. V. M\"akel\"a,$^{1}$
J. Ruohotie,$^{1}$\newauthor
C. P. Sishtla$^{2}$
and J. E. Soljento$^{1}$
\\
$^{1}$Department of Physics, University of Helsinki, PO Box 64, 00014 Helsinki, Finland\\
$^{2}$Department of Physics and Astronomy, Queen Mary University of London, London E1 4NS, UK
}

\date{Accepted 2025 September 24. Received 2025 September 23; in original form 2025 June 16}

\pubyear{2025}

\begin{document}
\label{firstpage}
\pagerange{\pageref{firstpage}--\pageref{lastpage}}
\maketitle

\begin{abstract}
Residual energy quantifies the difference in energy between velocity and magnetic field fluctuations in a plasma. Recent observational evidence highlights that fast-mode interplanetary shock waves have positive residual energy, in sharp contrast to the negative residual energy of the turbulence and magnetic structures that constitute the vast majority of fluctuation power in the solar wind at magnetohydrodynamic (MHD) inertial scales. In this work, we apply the Rankine-Hugoniot conditions to derive an equation for the residual energy of an MHD shock jump as a function of the shock angle, density compression ratio and Alfv\'en Mach number upstream of the shock. An equation for the cross helicity is similarly derived. The residual energy equation gives only positive values for super-Alfv\'enic (i.e. fast-mode) shocks. The residual energy and cross helicity of slow-mode shocks and tangential, contact and rotational discontinuities are also determined. A simplified form of the residual energy equation applicable to perpendicular shocks has been verified against residual energy values directly estimated from observations of 141 interplanetary shocks; the equation is found to match well with observations, particularly for shocks with higher density compression ratios and Mach numbers. The use of positive residual energy as a signature for fast-mode shock identification in spacecraft data is briefly considered, and insights from this work relating to compressive fluctuations more generally in the solar wind are discussed.
\end{abstract}


\section{Introduction}
\label{sec:intro}

Plasmas support waves, turbulence and structures that are collectively manifested as fluctuations of the velocity, magnetic field and various other parameters that determine the state of the plasma. In the solar wind, our most readily observable example of a collisionless plasma, the majority of fluctuation power at magnetohydrodynamic (MHD) scales is contained in Alfv\'enic modes \citep{BelcherDavis71}, with a minority compressive component \citep[typically 2--10 per cent of total power at 1~au; e.g.][]{Chen16} also present. The compressive component is mostly slow-mode-like with a much smaller fraction of power in the fast mode, these two modes being characterized by anti-correlations and correlations of density and magnetic field magnitude fluctuations, respectively \citep[e.g.][and references therein]{Verscharen17}. A striking demonstration of solar wind compressibility is the presence of shock waves, which develop from the steepening of compressive waves \citep[e.g.][]{Burgess95}. Shock waves in the heliosphere are usually fast-mode in nature and driven by transient structures \citep[i.e. coronal mass ejections and stream interaction regions;][]{Kilpua17,Richardson18} or form upstream of planetary magnetospheres \citep{Russell85,SpreiterStahara95}. 

In the solar wind, the spectrum of fluctuations at MHD scales smaller than the correlation length in the plasma is consistent with the presence of a turbulent cascade \citep[][]{BrunoCarbone16}. One widely observed feature of the solar wind predicted by Alfv\'enic turbulence theories and modelling is an excess of energy in magnetic field fluctuations, $E_b\sim\delta b^2$, relative to the energy in velocity fluctuations, $E_u\sim\delta u^2$. This residual energy is defined in its normalized form as $\sigma_\textrm{r} = (E_u - E_b)/(E_u + E_b)$, and is thus usually negative in the solar wind due to the $E_b$ excess; a related parameter is the Alfv\'en ratio, $r_\textrm{A}=E_u/E_b$, which is usually less than unity. Ideal Alfv\'en waves are in energy equipartition (i.e. $\sigma_\textrm{r} = 0$), and the non-equipartition in the solar wind is thought to arise from the fluctuations being non-linear and turbulent rather than purely linear and wave-like \citep[e.g.][]{Dorfman25}. Besides the action of Alfv\'enic turbulence, other possible sources of negative residual energy include intermittency (itself an inherent feature of turbulence) being greater in the magnetic field than velocity, and the advection of magnetic structures that are separate to the turbulence. \citet{Chen13} review a wide range of theories that have been proposed to account for the solar wind's negative residual energy.

Time series of solar wind data can be productively analysed using wavelet methods, which provide fluctuation spectral power as functions of frequency and time \citep{Torrence98}. Retention of the time domain, which is usually lost with traditional Fourier and structure-function analysis methods, is essential when performing the holistic analysis of strongly inhomogeneous intervals containing a range of different large-scale structures and solar wind types. Moreover, wavelet spectrograms of residual energy are now routinely presented in observational studies of solar wind fluctuations, with the analysed intervals sometimes encompassing interplanetary shocks \citep[][]{Zhao21b,Ruohotie22,Soljento23,Scolini24,Scolini25}. A feature of these shock-encompassing spectrograms that has not been previously commented upon is the positive residual energy signature that typically accompanies the shock wave, with this positive signature often standing out conspicuously from the negative residual energy spectrum of the ambient solar wind at MHD inertial scales. A motivation for this work has been to characterize the shock-associated residual energy that appears in these spectrograms. A second, broader motivation has been to consider the impact of compressive fluctuations more generally on measurements of solar wind residual energy.

It may be expected \textit{a priori} that fast-mode shocks are not in equipartition because they are intrinsically non-Alfv\'enic and may retain certain properties of their fast-wave precursors. For example, the MHD equations can be used to show that small-amplitude linear fast-mode waves with wavevectors perpendicular to the mean magnetic field are polarized approximately as $\delta u/v_\textrm{f}\sim \delta b/v_\textrm{A}$, where $v_\textrm{f}$ and $v_\textrm{A}$ are the fast-mode and Alfv\'en wave speeds, respectively \citep[e.g.][]{Zhao25b}; given that $v_\textrm{f}>v_\textrm{A}$, such waves must have $\sigma_\textrm{r}>0$. Using expressions obtained from the Rankine-Hugoniot relations, which capture the fully non-linear behaviour of shocks at any amplitude, large or small, we show that the corresponding residual energy of fast-mode shocks is also non-zero and positive. In Section~\ref{sec:equation}, an exact expression for the normalized residual energy of an idealized MHD shock as a function of the upstream Alfv\'en Mach number, density compression ratio and shock angle is derived. An equivalent expression for the normalized cross helicity, which measures the velocity--magnetic field correlation, is also derived; other types of MHD discontinuity are also examined in Section~\ref{sec:equation}. In Section~\ref{sec:observations}, the residual energy expression for the simplified case of a perpendicular shock is found to be in good agreement with observations of real shocks listed in the IP Shocks Database \citep{Kilpua15}, which applies a standardized analysis procedure to shocks observed by a range of spacecraft across the heliosphere. The value of using the positive residual energy signature of fast-mode shocks in wavelet spectrograms as a shock identification parameter is also briefly considered in Section~\ref{sec:observations}. Some of the wider implications of the results are discussed in Section~\ref{sec:discussion}. Finally, we emphasize that this study is concerned with the residual energy of the shock jump itself rather than the residual energy or Alfv\'en ratio of the waves, turbulence and structures that may be generated by or transmitted across the shock \citep[][]{Adhikari16,Pitna17,Borovsky20,Zank21,Nakanotani22,Kilpua25}; a comprehensive theoretical framework describing the shock-turbulence interaction and its effect on fluctuation properties nearby shocks was recently presented by \citet{Zhao25b}.

\section{Residual Energy Equation}
\label{sec:equation}

We now derive expressions for the residual energy and cross helicity of a shock in terms of key shock parameters, with the treatment of shocks following that of \citet{Burgess95}. The derivation is performed in the shock rest frame for a one-dimensional MHD shock conforming to the Rankine-Hugoniot jump conditions. Subscripts `u' and `d' are appended to parameters upstream and downstream of the shock, respectively, subscripts `{\it n}' and `{\it t}' refer to parameters normal and tangential to the shock surface, respectively, and the change across the shock in any parameter $X$ is denoted by $[X]=X_\textrm{u}-X_\textrm{d}$. The general form of the derivation is appropriate for a discontinuity that is compressive and that has a non-zero inflow velocity, both of which are required for a discontinuity to meet the definition of a shock. Moreover, the derivation concerns only the bulk transition across an idealized shock and not any other fluctuations (e.g. relating to turbulence) that may be present across or in the vicinity of the shock jump in more realistic situations.

The normalized residual energy of a shock jump is given by
\begin{equation}
\sigma_\textrm{r}=\frac{[\textit{\textbf{u}}]\!\cdot\![\textit{\textbf{u}}]-
[\textit{\textbf{b}}]\!\cdot\![\textit{\textbf{b}}]}
{[\textit{\textbf{u}}]\!\cdot\![\textit{\textbf{u}}]+
[\textit{\textbf{b}}]\!\cdot\![\textit{\textbf{b}}]}
\label{eqn:sigma_r1}
\end{equation}
and the cross helicity by
\begin{equation}
\sigma_\textrm{c}=\frac{2[\textit{\textbf{u}}]\!\cdot\![\textit{\textbf{b}}]}
{[\textit{\textbf{u}}]\!\cdot\![\textit{\textbf{u}}]+
[\textit{\textbf{b}}]\!\cdot\![\textit{\textbf{b}}]}
\label{eqn:sigma_c1}
\end{equation}
where $\textit{\textbf{u}}$ is the velocity, $\textit{\textbf{b}}=\textit{\textbf{B}}/\sqrt{\mu_0\rho}$ is the magnetic field expressed in Alfv\'en units and $\rho$ is the density. We begin by considering general expressions for the upstream and downstream magnetic fields,
\begin{subequations} \label{eq:Bs}
    \begin{align}
        \textit{\textbf{B}}_\textrm{u} &= \textit{\textbf{B}}_{\textrm{u}n} + \textit{\textbf{B}}_{\textrm{u}t} = s_nB_{\textrm{u}n}\hat{\textit{\textbf{n}}} + s_tB_{\textrm{u}t}\hat{\textit{\textbf{t}}} \label{eq:Bu} \\
        \textit{\textbf{B}}_\textrm{d}  &= \textit{\textbf{B}}_{\textrm{d}n} + \textit{\textbf{B}}_{\textrm{d}t} = s_nB_{\textrm{d}n}\hat{\textit{\textbf{n}}} + s_tB_{\textrm{d}t}\hat{\textit{\textbf{t}}}, \label{eq:Bd}
    \end{align}
\end{subequations}
such that
\begin{equation}
[\textit{\textbf{B}}] = s_n[B_n]\hat{\textit{\textbf{n}}}+s_t[B_t]\hat{\textit{\textbf{t}}},
\label{eqn:delB}
\end{equation}
where $\hat{\textit{\textbf{n}}}$ and $\hat{\textit{\textbf{t}}}$ are unit vectors normal and tangential to the shock, respectively, with $\hat{\textit{\textbf{n}}}$ pointing away from the shock in the upstream, where $B_{\textrm{u}n}=|\textit{\textbf{B}}_\textrm{u}\!\cdot\!\hat{\textit{\textbf{n}}}|$ etc. are the component magnitudes, and where $s_n=\pm 1$ and $s_t=\pm 1$ are the component signs. Thus $[\textit{\textbf{b}}]$ may be written as
\begin{equation}
[\textit{\textbf{b}}] = s_n\left(\frac{B_{\textrm{u}n}}{\sqrt{\mu_0\rho_\textrm{u}}}-\frac{B_{\textrm{d}n}}{\sqrt{\mu_0\rho_\textrm{d}}}\right)\hat{\textit{\textbf{n}}}+s_t\left(\frac{B_{\textrm{u}t}}{\sqrt{\mu_0\rho_\textrm{u}}}-\frac{B_{\textrm{d}t}}{\sqrt{\mu_0\rho_\textrm{d}}}\right)\hat{\textit{\textbf{t}}}.
\label{eqn:delb1}
\end{equation}
 A divergence-free magnetic field satisfying $\nabla\!\cdot\!\textit{\textbf{B}}=0$ imposes continuity of the normal component of the magnetic field (i.e. $[B_n]=0$), allowing us to write the magnitude of the normal component of $[\textit{\textbf{b}}]$ as
\begin{equation}
[b_n] = \frac{B_{\textrm{u}n}}{\sqrt{\mu_0\rho_\textrm{u}}} \left(1-\sqrt{\frac{\rho_\textrm{u}}{\rho_\textrm{d}}}\right).
\label{eqn:delbn1}
\end{equation}
We now introduce as variables the density compression ratio, $r=\rho_\textrm{d}/\rho_\textrm{u}$, the angle between the shock normal and upstream magnetic field, $\theta_{Bn}=\cos{\!^{-1}(B_{\textrm{u}n}/|\textit{\textbf{B}}_\textrm{u}|})$, and the upstream Alfv\'en Mach number, $M_\textrm{A}=u_{\textrm{u}n}\sqrt{\mu_0 \rho_\textrm{u}}/|\textit{\textbf{B}}_\textrm{u}|$; equation~(\ref{eqn:delbn1}) may be rewritten in terms of these variables as
\begin{equation}
[b_n] = \frac{u_{\textrm{u}n}}{M_\textrm{A}} \cos{\theta_{Bn}} \left( 1 - \frac{1}{\sqrt{r}} \right).
\label{eqn:delbn2}
\end{equation}
In a similar fashion, the magnitude of the tangential component in equation~(\ref{eqn:delb1}) may be written as
\begin{equation}
[b_t] = \frac{B_{\textrm{u}t}}{\sqrt{\mu_0\rho_\textrm{u}}} \left(1-\frac{B_{\textrm{d}t}}{B_{\textrm{u}t}}\sqrt{\frac{\rho_\textrm{u}}{\rho_\textrm{d}}}\right)
= \frac{u_{\textrm{u}n}}{M_\textrm{A}} \sin{\theta_{Bn}} \left( 1 - \frac{B_{\textrm{d}t}}{B_{\textrm{u}t}}\frac{1}{\sqrt{r}} \right),
\label{eqn:delbt1}
\end{equation}
noting that here $\theta_{Bn}=\sin{\!^{-1}(B_{\textrm{u}t}/|\textit{\textbf{B}}_\textrm{u}|})$. An expression for $B_{\textrm{d}t}/B_{\textrm{u}t}$ is required for further simplification of equation~(\ref{eqn:delbt1}). By eliminating $[\textit{\textbf{u}}_t]$ from the Rankine-Hugoniot jump conditions describing balance of the transverse momentum, 
\begin{equation}
\left[\rho u_n\textit{\textbf{u}}_t-\frac{B_n\textit{\textbf{B}}_t}{\mu_0}\right]=0, 
\label{eqn:tmomcon}
\end{equation}
and continuity of the tangential component of the electric field,
\begin{equation}
[u_n\textit{\textbf{B}}_t-B_n\textit{\textbf{u}}_t]=0, 
\label{eqn:Etancon}
\end{equation}
it may be shown that
\begin{equation}
\left[\left(1 - \frac{B_n^2}{\mu_0 \rho u_n^2} \right) u_n\textit{\textbf{B}}_t \right]=0.
\label{eqn:ut_removed}
\end{equation}
Furthermore, by noting that continuity of a non-zero mass flux through the shock (i.e. $[\rho u_n]=0$) allows us to write $u_{\textrm{d}n}=(1/r)u_{\textrm{u}n}$, by using the substitution $[\textit{\textbf{B}}_t]=s_t[B_t]\hat{\textit{\textbf{t}}}$ from equation~(\ref{eqn:delB}), and by recalling that $[B_n]=0$, it may be shown that a rearrangement of equation~(\ref{eqn:ut_removed}) gives
\begin{equation}
\frac{B_{\textrm{d}t}}{B_{\textrm{u}t}}=\chi r
\label{eqn:Bt_ratio}
\end{equation}
in terms of the shock parameters, where
\begin{equation}
\chi = \frac{M_\textrm{A}^2-\cos{\!^2\theta_{Bn}}}{M_\textrm{A}^2-r\cos{\!^2\theta_{Bn}}}.
\label{eqn:chi}
\end{equation}
Thus,
\begin{equation}
[b_t] = \frac{u_{\textrm{u}n}}{M_\textrm{A}} \sin{\theta_{Bn}} \left( 1 - \chi\sqrt{r} \right)
\label{eqn:delbt2}
\end{equation}
and equation~(\ref{eqn:delb1}) may be written as
\begin{equation}
[\textit{\textbf{b}}] = \frac{s_nu_{\textrm{u}n}}{M_\textrm{A}} \cos{\theta_{Bn}} \left( 1 - \frac{1}{\sqrt{r}} \right) \hat{\textit{\textbf{n}}}+\frac{s_tu_{\textrm{u}n}}{M_\textrm{A}} \sin{\theta_{Bn}} \left( 1 - \chi\sqrt{r} \right)\hat{\textit{\textbf{t}}}.
\label{eqn:delb2}
\end{equation}
Turning to the velocity, we begin as for the magnetic field with generalized upstream and downstream expressions,
\begin{subequations} \label{eq:us}
    \begin{align}
        \textit{\textbf{u}}_\textrm{u} &= \textit{\textbf{u}}_{\textrm{u}n} + \textit{\textbf{u}}_{\textrm{u}t} = -u_{\textrm{u}n}\hat{\textit{\textbf{n}}} + \textit{\textbf{u}}_{\textrm{u}t} \label{eq:uu} \\
        \textit{\textbf{u}}_\textrm{d}  &= \textit{\textbf{u}}_{\textrm{d}n} + \textit{\textbf{u}}_{\textrm{d}t} = -u_{\textrm{d}n}\hat{\textit{\textbf{n}}} + \textit{\textbf{u}}_{\textrm{d}t}, \label{eq:ud}
    \end{align}
\end{subequations}
where the tangential vectors have been left in the more general form. The negative signs of the normal components reflect the requirement of the flow to be towards the shock in the upstream and away from the shock in the downstream. Thus, the velocity jump is given by 
\begin{equation}
[\textit{\textbf{u}}] = -\left(u_{\textrm{u}n}-u_{\textrm{d}n}\right) \hat{\textit{\textbf{n}}} + [\textit{\textbf{u}}_t].
\label{eqn:delu1}
\end{equation}
The magnitude of the normal component can be directly written as
\begin{equation}
[u_n]=u_{\textrm{u}n}\left( 1 - \frac{1}{r}\right)
\label{eqn:delun}
\end{equation}
in terms of $r$. Expansion of equation~(\ref{eqn:tmomcon}) with $[\textit{\textbf{B}}_t]=s_t[B_t]\hat{\textit{\textbf{t}}}$ gives an expression for the tangential velocity jump,
\begin{equation}
[\textit{\textbf{u}}_t] = \frac{s_t B_{\textrm{u}n}B_{\textrm{u}t}}{\mu_0\rho_\textrm{u}u_{\textrm{u}n}} \left(1-\frac{B_{\textrm{d}t}}{B_{\textrm{u}t}}\right) \hat{\textit{\textbf{t}}},
\label{eqn:delut1}
\end{equation}
which may be written as
\begin{equation}
[\textit{\textbf{u}}_t] = \frac{s_t u_{\textrm{u}n}}{M_\textrm{A}^2} \sin{\theta_{Bn}} \cos{\theta_{Bn}} \left( 1 - \chi r \right) \hat{\textit{\textbf{t}}}.
\label{eqn:delut2}
\end{equation}
Thus, equation~(\ref{eqn:delu1}) may be written as
\begin{equation}
[\textit{\textbf{u}}] = -u_{\textrm{u}n} \left( 1 - \frac{1}{r} \right) \hat{\textit{\textbf{n}}}+\frac{s_t u_{\textrm{u}n}}{M_\textrm{A}^2} \sin{\theta_{Bn}} \cos{\theta_{Bn}} \left( 1 - \chi r \right)\hat{\textit{\textbf{t}}}.
\label{eqn:delu2}
\end{equation}
Via substitution of equations~(\ref{eqn:delb2}) and (\ref{eqn:delu2}) into equations~(\ref{eqn:sigma_r1}) and (\ref{eqn:sigma_c1}), we are now able to define in terms of $r$, $M_\textrm{A}$, $\theta_{Bn}$ and $s_n$ the residual energy,
\begin{equation}
\sigma_{\textrm{r}} = \frac{\left(1-\frac{1}{r} \right)^2 + C_1\left(1-\chi r \right)^2 - C_2\left(1-\frac{1}{\sqrt{r}} \right)^2 - C_3\left(1-\chi \sqrt{r} \right)^2}{\left(1-\frac{1}{r} \right)^2 + C_1\left(1-\chi r \right)^2 + C_2\left(1-\frac{1}{\sqrt{r}} \right)^2 + C_3\left(1-\chi \sqrt{r} \right)^2},
\label{eqn:sigma_r2}
\end{equation}
and the cross helicity,
\begin{equation}
\sigma_{\textrm{c}} = \frac{-s_nC_4\left( 1-\frac{1}{\sqrt{r}}\right)\left( 1-\frac{1}{r}\right)+C_5\left(1-\chi\sqrt{r}\right)\left(1-\chi r\right)}{\left(1-\frac{1}{r} \right)^2 + C_1\left(1-\chi r \right)^2 + C_2\left(1-\frac{1}{\sqrt{r}} \right)^2 + C_3\left(1-\chi \sqrt{r} \right)^2},
\label{eqn:sigma_c2}
\end{equation}
where
\begin{subequations} \label{eq:constants}
    \begin{align}
        C_1 &= \frac{1}{M_\textrm{A}^4}\sin{\!^2\theta_{Bn}}\cos{\!^2\theta_{Bn}}  \label{eq:C1} \\
        C_2 &= \frac{1}{M_\textrm{A}^2}\cos{\!^2\theta_{Bn}}  \label{eq:C2} \\
        C_3 &= \frac{1}{M_\textrm{A}^2}\sin{\!^2\theta_{Bn}}  \label{eq:C3} \\
        C_4 &= \frac{2}{M_\textrm{A}}\cos{\theta_{Bn}}  \label{eq:C4} \\
        C_5 &= \frac{2}{M_\textrm{A}^3}\sin{\!^2\theta_{Bn}}\cos{\theta_{Bn}}.  \label{eq:C5}
    \end{align}
\end{subequations}
The cross helicity is dependent on $s_n$ but not $s_t$, while the residual energy is dependent on neither. The Alfv\'en ratio may similarly be defined as
\begin{equation} \label{eqn:rA}
r_{\textrm{A}}=\frac{[\textit{\textbf{u}}]\!\cdot\![\textit{\textbf{u}}]}{[\textit{\textbf{b}}]\!\cdot\![\textit{\textbf{b}}]} = \frac{\left(1-\frac{1}{r} \right)^2 + C_1\left(1-\chi r \right)^2}{C_2\left(1-\frac{1}{\sqrt{r}} \right)^2 + C_3\left(1-\chi \sqrt{r} \right)^2}.
\end{equation}

\begin{figure*}
	\includegraphics[width=2\columnwidth]{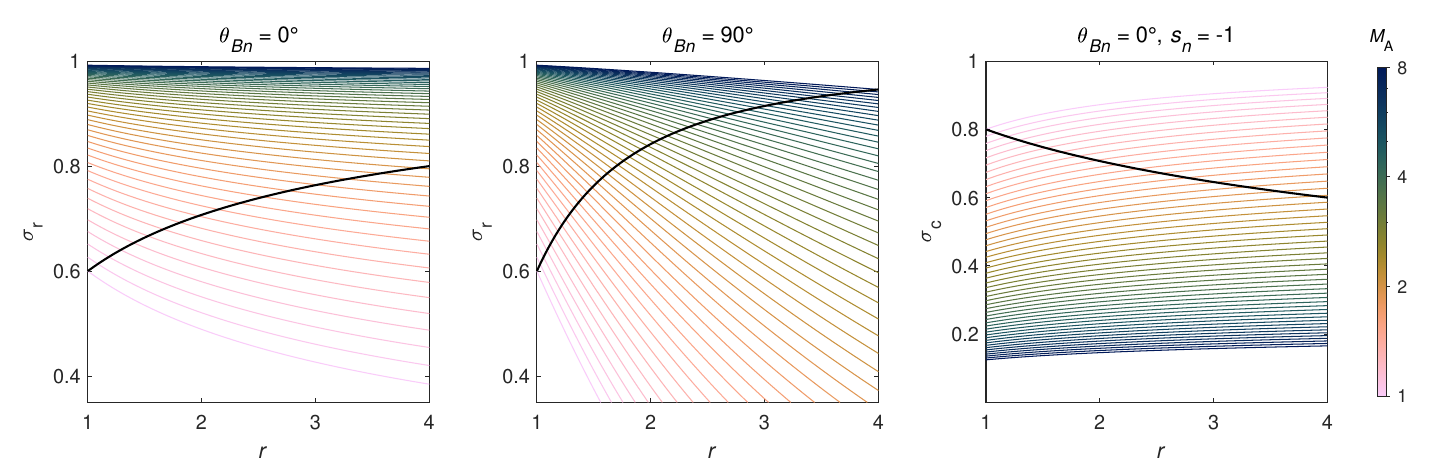}
    \caption{Residual energy of the parallel shock (left-hand panel) and perpendicular shock (middle panel), and cross helicity of the parallel shock (right-hand panel) with the upstream magnetic field pointing towards the shock surface, as functions of the density compression ratio, $r$, and upstream Alfv\'en Mach number, $M_\textrm{A}$. Super-Alfv\'enic shock solutions lie above the black curves in the residual energy panels and below the curve in the cross helicity panel.}
    \label{fig:fixed_theta}
\end{figure*}

Equations~(\ref{eqn:sigma_r2}), (\ref{eqn:sigma_c2}) and (\ref{eqn:rA}) simplify substantially for the special cases of the exactly parallel shock ($\theta_{Bn}=0^{\circ}$) and exactly perpendicular shock ($\theta_{Bn}=90^{\circ}$). The residual energy given by equation~(\ref{eqn:sigma_r2}) simplifies in the parallel case to
\begin{equation}
\sigma_{\textrm{r}\parallel}=\frac{M_\textrm{A}^2\left(1+\sqrt{r}\right)^2-r}{M_\textrm{A}^2\left(1+\sqrt{r}\right)^2+r}
\label{eqn:sigma_r_par}
\end{equation}
and in the perpendicular case to
\begin{equation}
\sigma_{\textrm{r}\bot}=\frac{M_\textrm{A}^2\left(1+\sqrt{r}\right)^2-r^2}{M_\textrm{A}^2\left(1+\sqrt{r}\right)^2+r^2}.
\label{eqn:sigma_r_perp}
\end{equation}
Likewise, equation~(\ref{eqn:sigma_c2}) for the cross helicity simplifies in the parallel case to
\begin{equation}
\sigma_{\textrm{c}\parallel}=-2 s_n\frac{M_\textrm{A}\left(r+\sqrt{r}\right)}{M_\textrm{A}^2\left(1+\sqrt{r}\right)^2+r}
\label{eqn:sigma_c_par}
\end{equation}
and in the perpendicular case to
\begin{equation}
\sigma_{\textrm{c}\bot}=0.
\label{eqn:sigma_c_perp}
\end{equation}
Finally, the Alfv\'en ratio of the parallel shock is given by
\begin{equation}
r_{\textrm{A}\parallel}=\frac{M_\textrm{A}^2\left(1+\sqrt{r}\right)^2}{r}
\label{eqn:rA_par}
\end{equation}
and the perpendicular shock by
\begin{equation}
r_{\textrm{A}\bot}=\frac{M_\textrm{A}^2\left(1+\sqrt{r}\right)^2}{r^2}.
\label{eqn:rA_perp}
\end{equation}

\subsection{Fast-mode shock solutions}
\label{sec:fast-mode}

Fig.~\ref{fig:fixed_theta} shows contour plots of equations~(\ref{eqn:sigma_r_par}) -- (\ref{eqn:sigma_c_par}) for $1 < r \leq 4$ and $1 \leq M_\textrm{A} \leq 8$, a suitable parameter space for examining fast-mode shock solutions. It can be seen that $\sigma_{\textrm{r}\parallel}$ and $\sigma_{\textrm{r}\bot}$ increase with $M_\textrm{A}$ and decrease with $r$, with $\sigma_{\textrm{r}\bot}$ displaying the stronger $r$ dependence. The reverse trends are seen in $\sigma_{\textrm{c}\parallel}$, here shown for $s_n=-1$. For all three parameters, the $r$ dependence diminishes with increasing $M_\textrm{A}$. 

Fast-mode shocks are super-Alfv\'enic, that is to say, they have flow speeds in the shock rest frame that exceed the Alfv\'en speed in both the upstream and downstream, with the flow speed in the downstream dropping below the fast-mode speed only; by requiring the downstream Alfv\'en Mach number to exceed unity (i.e. $u_{\textrm{d}n}\sqrt{\mu_0 \rho_\textrm{d}}/|\textit{\textbf{B}}_\textrm{d}|>1$), it may be shown that super-Alfv\'enic shocks have $M_\textrm{A}>\sqrt{r}$ in the exactly parallel case and $M_\textrm{A}>\sqrt{r^3}$ in the exactly perpendicular case. The black curves in Fig.~\ref{fig:fixed_theta} mark the resulting boundaries between super-Alfv\'enic and sub-Alfv\'enic solutions to equations~(\ref{eqn:sigma_r_par}) -- (\ref{eqn:sigma_c_par}); super-Alfv\'enic solutions lie above the curves in the residual energy panels and below the curve in the cross helicity panel. Only positive values of residual energy are found in the super-Alfv\'enic parameter space.

\begin{figure*}
	\includegraphics[width=2\columnwidth]{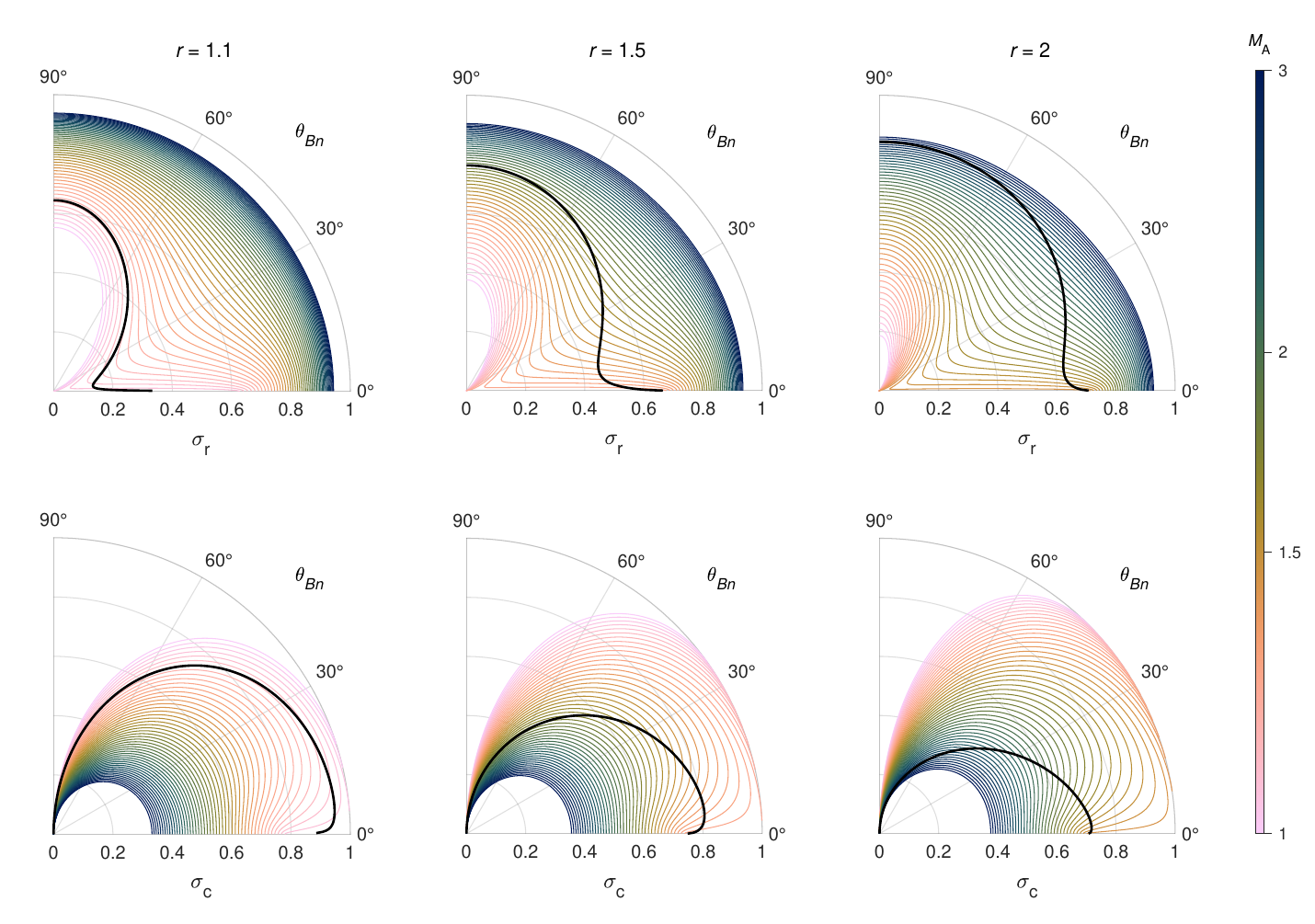}
    \caption{Polar plots of residual energy (top panels) and cross helicity (bottom panels, for $s_n=-1$) as functions of the shock angle, $\theta_{Bn}$, and upstream Alfv\'en Mach number, $M_\textrm{A}$, shown for three values of the density compression ratio, $r$. Super-Alfv\'enic shock solutions lie above and below the black curves in the residual energy and cross helicity panels, respectively.}
    \label{fig:polar_theta}
\end{figure*}

Fig.~\ref{fig:polar_theta} shows the full $\theta_{Bn}$ dependence of $\sigma_{\textrm{r}}$ and $\sigma_{\textrm{c}}$ with polar contour plots of equations~(\ref{eqn:sigma_r2}) and (\ref{eqn:sigma_c2}), respectively, for ${1 \leq M_\textrm{A} \leq 3}$ and $r=$ 1.1, 1.5 and 2. A contour is only displayed at a given angle if $\chi r>1$, which, as indicated by equation~(\ref{eqn:Bt_ratio}), corresponds to the magnetic field bending away from the shock normal in the downstream, as must happen for an oblique fast-mode shock. Black curves demarcate sub- and super-Alfv\'enic solutions as in Fig.~\ref{fig:fixed_theta}, with the super-Alfv\'enic region above (below) the curves in the $\sigma_{\textrm{r}}$ ($\sigma_{\textrm{c}}$) panels. Here, the more general inequality satisfied by a super-Alfv\'enic shock,
\begin{equation}
M_\textrm{A}>\sqrt{r\cos{\!^2\theta_{Bn}}+r^3\chi^2\sin{\!^2\theta_{Bn}}},
\label{eqn:max_MA}
\end{equation}
has been solved numerically to find $M_\textrm{A}$ and thus, via substitution into equations~(\ref{eqn:sigma_r2}) and (\ref{eqn:sigma_c2}), the super-Alfv\'enic limits of $\sigma_{\textrm{r}}$ and $\sigma_{\textrm{c}}$. Various trends in the figure are evident, including: increasingly small super-Alfv\'enic regions with increasing $r$; contours at fixed $M_\textrm{A}$ lying at lower (higher) $\sigma_{\textrm{r}}$ ($\sigma_{\textrm{c}}$) values with increasing $r$; and $\sigma_{\textrm{r}}$ contours dipping at oblique $\theta_{Bn}$ values, with this effect most significant at quasi-parallel $\theta_{Bn}$.

\subsection{Slow-mode shock solutions}
\label{sec:slow-mode}

\begin{figure}
	\includegraphics[width=\columnwidth]{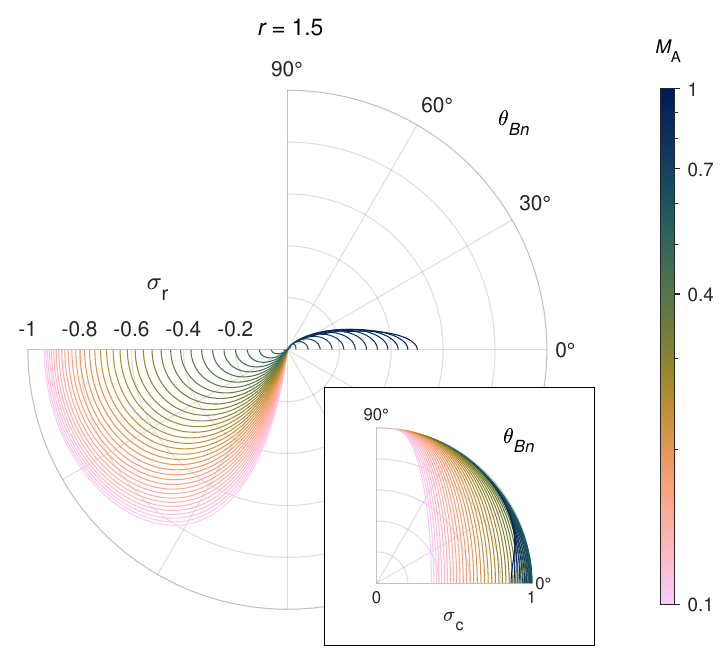}
    \caption{Polar plots of residual energy (main panel) and cross helicity (inset panel, for $s_n=-1$) for slow-mode shocks at $0.1<M_\textrm{A}<1$ and $r=1.5$. Values in the bottom-left quadrant of the residual energy panel are negative.}
    \label{fig:slow_shocks}
\end{figure}

Slow shock solutions have $M_\textrm{A}\lesssim1$ \citep[e.g.][]{BurgessScholer15} and $0<\chi r<1$, the latter condition corresponding to a magnetic field that bends towards the shock normal (and thus falls in magnitude) in the downstream. Solutions satisfying these conditions are shown in Fig.~\ref{fig:slow_shocks} for $r=1.5$. Negative values of $\sigma_{\textrm{r}}$ are projected to the bottom-left quadrant in the polar plot representation. It can be seen that $\sigma_{\textrm{r}}$ is negative at $M_\textrm{A}<0.5$ for this particular $r$ value. The $r$ dependence of the solutions in the negative-$\sigma_{\textrm{r}}$ quadrant is weak, and weakens with decreasing $M_\textrm{A}$. As $M_\textrm{A}$ falls increasingly below 0.5, $\sigma_{\textrm{r}}$ becomes more negative with solutions spanning a wider $\theta_{Bn}$ range that asymptotically broadens towards $\theta_{Bn}=90^{\circ}$. Exactly perpendicular slow-mode shocks are unable to form given that slow-mode waves cannot propagate perpendicularly to the magnetic field, hence the absence of solutions at $\theta_{Bn}=90^{\circ}$. Cross helicity falls from approximately unity at parallel and oblique $\theta_{Bn}$ angles as $M_\textrm{A}$ falls from 0.5, with the fall-off being stronger at more parallel $\theta_{Bn}$.

\subsection{Tangential, contact and rotational discontinuities}
\label{sec:other_disc}

In the limit of $M_\textrm{A}=0$ and $\theta_{Bn}$ tending to 90$^{\circ}$, the slow-mode shock becomes a non-propagating tangential discontinuity and $\sigma_{\textrm{r}}=-1$, independently of $r$. The contact discontinuity, which has $M_\textrm{A}=0$ and $\theta_{Bn}\neq90^{\circ}$, similarly has $\sigma_{\textrm{r}}=-1$. Both discontinuities have $\sigma_{\textrm{c}}=0$ given that $\sigma_{\textrm{r}}^2+\sigma_{\textrm{c}}^2\leq1$. Solutions with $M_\textrm{A}=0$ describe discontinuities that have no inward flow from the upstream (i.e. ${u_{\textrm{u}n}=0}$) and are thus not shock-like. 

The final canonical MHD discontinuity to be considered is the rotational discontinuity, which is incompressible (i.e. $r=1$), has a magnetic field that rotates $180^{\circ}$ in the plane of the discontinuity (i.e. $\textit{\textbf{b}}_{\textrm{u}t}=-\textit{\textbf{b}}_{\textrm{d}t}$), and has a magnetic field that is correlated or anti-correlated with the velocity (i.e. $\textit{\textbf{u}}=\pm \textit{\textbf{b}}$). It may thus be shown that $[\textit{\textbf{u}}_{n}]=[\textit{\textbf{b}}_{n}]=0$, $[\textit{\textbf{u}}]=2\textit{\textbf{u}}_{\textrm{u}t}$ and $[\textit{\textbf{b}}]=2\textit{\textbf{b}}_{\textrm{u}t}=\pm2\textit{\textbf{u}}_{\textrm{u}t}$; with substitution into equations~(\ref{eqn:sigma_r1}) and (\ref{eqn:sigma_c1}), we find that $\sigma_{\textrm{r}}=0$ and $\sigma_{\textrm{c}}=\pm1$, highlighting the fact that the rotational discontinuity closely resembles an ideal Alfv\'en wave. 

\section{Shock Observations}
\label{sec:observations}

\subsection{Comparison to theory}
\label{sec:verification}

\begin{figure*}
    \includegraphics[width=2.1\columnwidth]{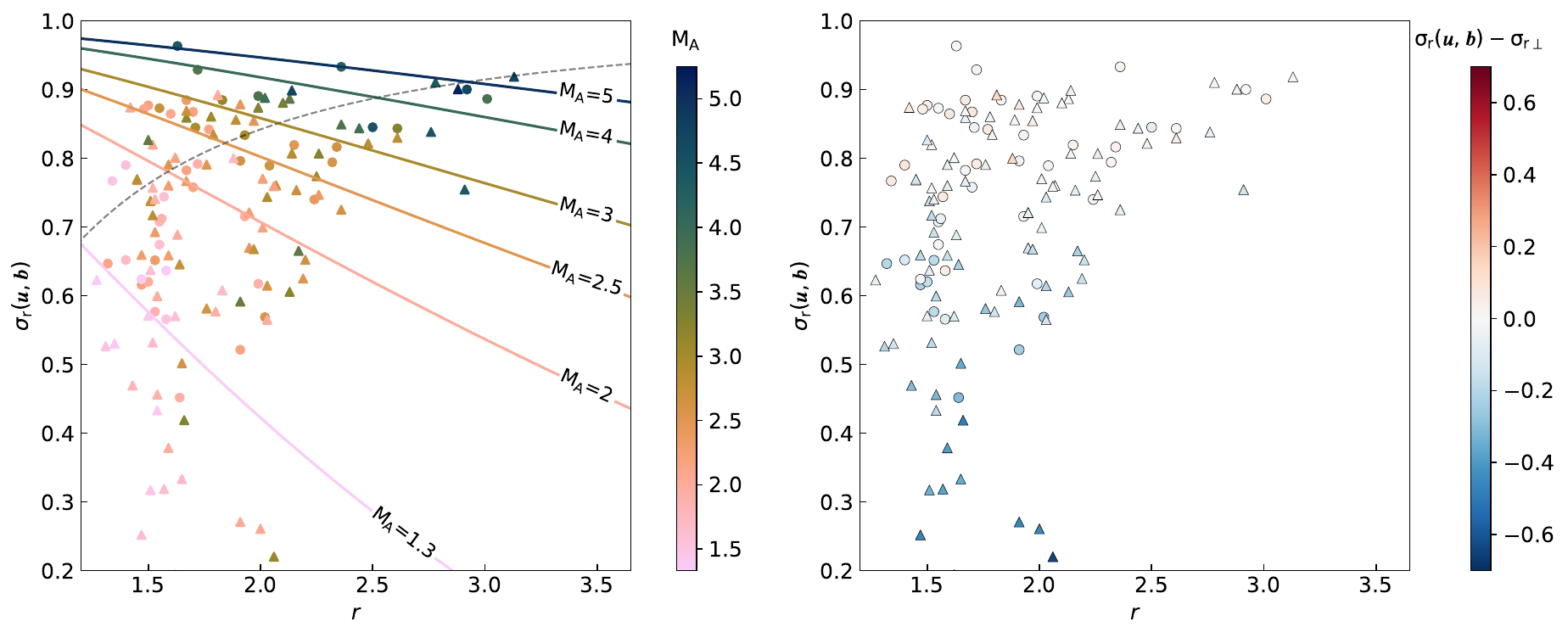}
    \caption{Residual energy of selected shocks from the IP Shocks Database versus $r$, with points colour-coded to $M_\textrm{A}$ (left-hand panel). Contours show the expected trends given by equation~(\ref{eqn:sigma_r_perp}) for perpendicular shocks. Points are also colour-coded to their deviation from the equation~(\ref{eqn:sigma_r_perp}) predictions (right-hand panel). Triangles and circles indicate shocks with $45^{\circ}<\theta_{Bn}<80^{\circ}$ and $\theta_{Bn}\geq80^{\circ}$, respectively.}
    \label{fig:verification_plot}
\end{figure*}

Equations derived in Section~\ref{sec:equation} give the residual energy of an idealized MHD shock in terms of $r$, $M_\textrm{A}$ and $\theta_{Bn}$. To test this dependence on shock parameters with observations of real shocks observed in the solar wind, the residual energies of fast-mode shocks listed in the IP Shocks Database have been directly calculated with equation~(\ref{eqn:sigma_r1}) using the upstream and downstream averages of the bulk proton velocity, \textit{\textbf{u}}, magnetic field, \textit{\textbf{B}}, and proton number density, $n_\textrm{p}$, provided in the database. The upstream and downstream averaging windows used in the IP Shocks methodology are each 8~min in duration, with the windows starting 9~min before and 2~min after the shock time, respectively. The 3-min cut-out excluding the shock discontinuity ensures that upstream and downstream conditions are not inadvertently mixed in the same window. The 8~min window duration was identified during the development of the IPShocks Database as being generally long enough to obtain accurate estimates of bulk upstream and downstream conditions without being so long that bulk variations unrelated to the shock begin to affect the averaging \citep{Kilpua15}.

The calculated $\sigma_{\textrm{r}}$ values have been plotted versus the relevant shock parameters available from the database and compared to the corresponding ideal values. In practice, the nature of the shocks available in the IP Shocks Database limits the scope for testing the theory presented in Section~\ref{sec:equation}. Given that most of the shocks were observed by spacecraft near 1~au, they are mostly quasi-perpendicular ($\theta_{Bn}>45^{\circ}$) rather than quasi-parallel ($\theta_{Bn}<45^{\circ}$). Exactly parallel shocks have $[B]=0$ and are thus excluded by the $B_\textrm{d}/B_\textrm{u}>1.2$ criterion required for a shock to be included in the database. Moreover, the $\theta_{Bn}$ estimates listed in the database carry significant uncertainties that preclude a detailed examination of the $\theta_{Bn}$ dependence predicted by equation~(\ref{eqn:sigma_r2}). Our efforts have therefore focussed on verifying equation~(\ref{eqn:sigma_r_perp}), which gives the residual energy of perpendicular shocks.

The left-hand panel of Fig.~\ref{fig:verification_plot} shows the residual energy of selected shocks from the IP Shocks Database as a function of $r$ and $M_\textrm{A}$, with the contours of constant $M_\textrm{A}$ given by equation~(\ref{eqn:sigma_r_perp}). This panel is directly comparable to the middle panel of Fig.\ref{fig:fixed_theta}. Residual energy values were calculated with mass density $\rho=7m_\textrm{p}n_\textrm{p}/6$, where $m_\textrm{p}$ is the proton mass and the $7/6$ factor arises from the assumption that 4 per cent of solar wind ions are alpha particles. The selected shocks were required to have $B_\textrm{d}/B_\textrm{u}$ within 5 per cent the value of the $r=\rho_\textrm{d}/\rho_\textrm{u}$ compression ratio. This criterion, based on the $r=B_\textrm{d}/B_\textrm{u}$ relation that holds for perpendicular shocks, was used to select shocks rather than setting a $\theta_{Bn}$ threshold. However, the $\theta_{Bn}$ estimates have also been taken into consideration. The 141 shocks included in this analysis that satisfy the $0.95<rB_u/B_d <1.05$ condition have $1 \leq r \leq 4$ and $1.3\leq M_\textrm{A}\leq 5.3$, with all having a nominal $\theta_{Bn}> 45^{\circ}$ (i.e. all being at least nominally quasi-perpendicular) and 48 having $\theta_{Bn}> 80^{\circ}$.

There is broad agreement between the trends in the calculated $\sigma_\textrm{r}$ values in Fig.~\ref{fig:verification_plot} and the equation~(\ref{eqn:sigma_r_perp}) contours. It can be seen that $\sigma_\textrm{r}$ increases with $M_\textrm{A}$ and decreases with $r$, with the $r$ dependence weakening as $M_\textrm{A}$ increases. Almost all of the $\sigma_\textrm{r}$ values are positive; six of the 141 shocks have values less than the $\sigma_\textrm{r} = 0.2$ cut-off in the figure (including two with negative values), this cut-off chosen to better display the parameter space where most data points are found. Values cluster around the ideal super-Alfv\'enic limit as shown by the dashed curve. While there is general agreement, the calculated $\sigma_\textrm{r}$ values show some significant scatter that is more commonly tilted towards lower values relative to the predicted contours, particularly for the shocks at low $r$ and $M_\textrm{A}$; this trend is highlighted in the right-hand panel of Fig.~\ref{fig:verification_plot}, in which points are colour-coded to their deviation from the predicted values. The trend could partly be explained by quasi-perpendicular shocks at low $r$ and $M_\textrm{A}$ having lower $\sigma_\textrm{r}$ than the exactly perpendicular case, as can be seen from the dipping contours in Fig.~\ref{fig:polar_theta}. For example, most of the low-$r$, low-$M_\textrm{A}$ shocks with $\sigma_\textrm{r}$ falling below the minimum $M_\textrm{A}=1.3$ contour in Fig.~\ref{fig:verification_plot} have more quasi-perpendicular $\theta_{Bn}$ angles (triangles: $45^{\circ}<\theta_{Bn}<80^{\circ}$). However, there is no significant difference in scatter between these nominally quasi-perpendicular shocks and the more perpendicular examples (circles: $\theta_{Bn}\geq80^{\circ}$) at higher $r$ and $M_\textrm{A}$. An Alfv\'enic component present in the ambient solar wind and contributing to the total $\textit{\textbf{u}}$ and $\textit{\textbf{b}}$ across the shock jump could, for example, also act to reduce the residual energy relative to the predicted contours. Such an Alfv\'enic component would be relatively more significant for weaker shocks at low $r$ and $M_\textrm{A}$.

\subsection{Wavelet spectrograms}
\label{sec:wavelets}

\begin{figure}
	\includegraphics[width=1.01\columnwidth]{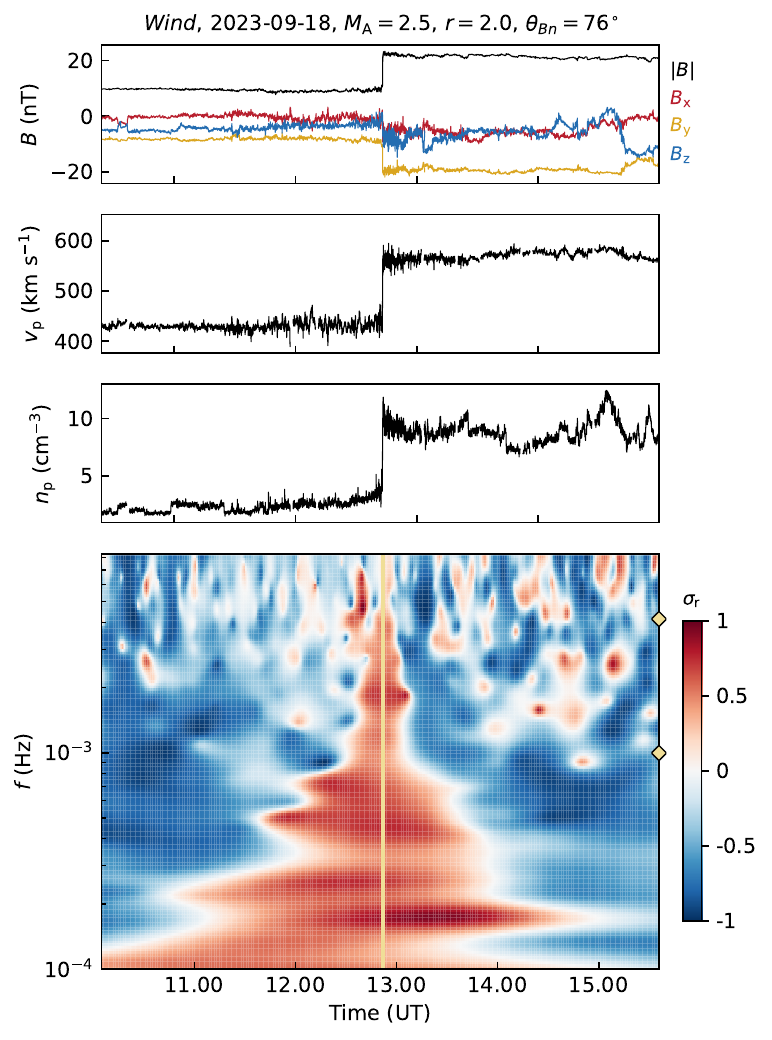}
    \caption{An example shock observed by the \textit{Wind} spacecraft. From top to bottom, the panels show the magnetic field magnitude and components in GSE coordinates, bulk proton speed, proton number density and wavelet spectrogram of the residual energy. The vertical line marks the shock time in the spectrogram, and the diamonds on the right-hand side of the panel mark the frequency range used for the analysis presented in Fig.~\ref{fig:wavelet_search}.}
    \label{fig:wavelet_plot}
\end{figure}

We now consider the signatures of shocks in wavelet spectrograms of residual energy as first discussed in Section~\ref{sec:intro}. Fig.~\ref{fig:wavelet_plot} displays observations made by the \textit{Wind} spacecraft of an interplanetary shock at 1~au, with magnetic field data from MFI \citep{Lepping95} and ion moments from 3DP/PESA-L \citep{Lin95}. Magnetic field components in the figure are shown in geocentric-solar-ecliptic (GSE) coordinates. The discontinuous increase in magnetic field strength, proton bulk speed and proton number density with time that characterizes a fast forward shock in the spacecraft reference frame is evident in the figure, with `forward' indicating an anti-sunward sense of propagation (cf. the discontinuous decrease in speed of a sunward-propagating `reverse' shock). The bottom panel of Fig.~\ref{fig:wavelet_plot} shows the residual energy spectrogram calculated with a Morlet wavelet such that
\begin{equation}
\sigma_{\textrm{r}}=\frac{\sum_i|\mathcal{W}_{u_i}|^2 - \sum_i|\mathcal{W}_{b_i}|^2}{\sum_i|\mathcal{W}_{u_i}|^2 + \sum_i|\mathcal{W}_{b_i}|^2},
\label{eqn:sigma_r_wav}
\end{equation}
where $\mathcal{W}_{u_i}=\mathcal{W}_{u_i}(f,t)$ and $\mathcal{W}_{b_i}=\mathcal{W}_{b_i}(f,t)$ are the wavelet transforms of the GSE components $i=x,y,z$ of the proton bulk velocity $u_i(t)$ and Alfv\'en-normalized magnetic field $b_i(t)=B_i(t)/\sqrt{\mu_0\rho(t)}$, respectively, and $\rho(t)=7m_\textrm{p}n_\textrm{p}(t)/6$. The spacecraft-frame frequency, $f$, is related to the wavelet timescale, $s$, as $f=1/s$. Fig.~\ref{fig:wavelet_plot} displays a segment of a spectrogram calculated over a longer time interval than that shown, with this segment falling entirely within the cone of influence of the longer spectrogram. The positive $\sigma_{\textrm{r}}$ centred on the shock time clearly stands out in the spectrogram against the predominantly negative $\sigma_{\textrm{r}}$ of the ambient solar wind, with the large-amplitude, step-like shock jump giving a signature that extends across the whole of the frequency range shown. The residual energy of an ideal shock is scale-independent; while showing some variability, the residual energy of the shock displayed in Fig.~\ref{fig:wavelet_plot} does not change in value significantly or systematically with frequency, in approximate agreement with the ideal case. The temporal width of the positive $\sigma_{\textrm{r}}$ signature scales approximately with the inverse of frequency, i.e. with $s$, consistent with the shock governing the residual energy of points in the spectrogram that have the shock jump within their cones of influence. Also present at $f\gtrsim1$~mHz in the spectrogram are localized patches of positive $\sigma_{\textrm{r}}$ away from the shock, which may be the signatures of non-shock compressions; since they occur at relatively high frequencies lying outside of the analysed range, the somewhat larger-amplitude density fluctuations visible for $\sim$30~min immediately downstream of the shock have no discernible signature in the $\sigma_{\textrm{r}}$ spectrogram.

\begin{figure}
	\includegraphics[width=1.03\columnwidth]{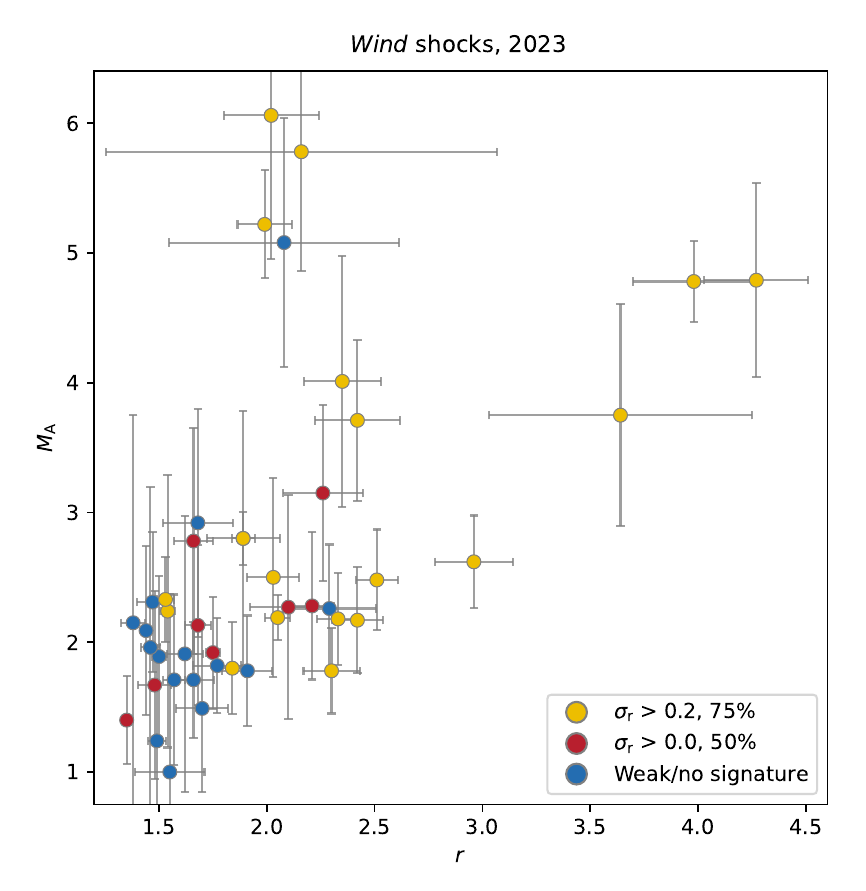}
    \caption{Alfv\'en Mach number versus compression ratio for 45 shocks observed by the \textit{Wind} spacecraft in 2023, with points colour-coded according to their residual energy signatures in wavelet spectrograms. Uncertainty values indicated by the error bars are taken from the IP Shocks Database.}
    \label{fig:wavelet_search}
\end{figure}

Fig.~\ref{fig:wavelet_search} shows $M_\textrm{A}$ versus $r$ for 45 shocks observed by \textit{Wind} during calendar year 2023, with points colour-coded to a categorization of their $\sigma_{\textrm{r}}$ signatures in wavelet spectrograms. These shocks were recently added to the IP Shocks Database. For each shock, $\sigma_{\textrm{r}}$ has been examined across the frequency band 1--4.17~mHz in spectrograms calculated with data at 1-min resolution. The upper limit of the band was chosen to be half the Nyquist frequency and the lower limit chosen to generally exclude any low-frequency bulk variations in the solar wind that might obscure the spectral signature of the shock. Note that Fig.~\ref{fig:wavelet_plot} presents an example where such bulk variations are not significant over the time period shown, such that the positive $\sigma_{\textrm{r}}$ of the shock extends to much lower frequencies than the 1~mHz band limit. While having no intrinsic significance relating to the physical properties of shock waves, the 1--4.17~mHz band falls within the inertial range of MHD turbulence at 1~au \citep[e.g.][]{Chen13}.

Residual energy values from the wavelet spectrogram for each shock have been sampled at 52 frequencies, evenly spaced over the logarithmically scaled 1--4.17~mHz band, with the sample taken at the observation time of the shock jump. Of the 45 shocks analysed: 20 shocks (yellow points in Fig.~\ref{fig:wavelet_search}) had (i) at least 75 per cent of the sampled $\sigma_{\textrm{r}}$ values exceeding 0.2 across the frequency band; nine shocks (red points) not meeting the case (i) criteria had (ii) at least 50 per cent of sampled values exceeding zero across the band; and 16 shocks (blue points) met neither the case (i) or (ii) criteria. It can be seen that the case (i) shocks, which had more sharply defined $\sigma_{\textrm{r}}$ signatures, tended to have higher $r$ and $M_\textrm{A}$. In contrast, the weaker shocks with lower $r$ and $M_\textrm{A}$ that met neither the case (i) nor (ii) criteria typically had some positive $\sigma_{\textrm{r}}$ signature that occupied less than 50 per cent of the frequency band. The continuum of positive $\sigma_{\textrm{r}}$ in the spectrograms appeared to be broken up in some cases by non-shock-related features spanning the shock jump that varied at timescales corresponding to the investigated frequency band (cf. the more ideal example shown in Fig.~\ref{fig:wavelet_plot}). The $\sigma_{\textrm{r}}$ signature may also have been less distinct for more parallel shocks given that they often show a less sharply defined jump in time series data, although we have not explored any quantitative $\theta_{Bn}$ dependence in this preliminary exercise. A shock-identification scheme based on a simple categorization of $\sigma_{\textrm{r}}$ signatures as described above -- i.e. positive, broadband $\sigma_{\textrm{r}}$ -- could be used to identify stronger shocks using wavelet spectrograms, but would likely miss weaker shocks if used in isolation. However, $\sigma_{\textrm{r}}$ spectrograms could provide useful additional information that complements existing identification schemes.

\section{Discussion}
\label{sec:discussion}

The concept of residual energy takes the equipartitioned MHD Alfv\'en wave as a zero reference point and associates negative (positive) values with an excess of magnetic field (velocity) fluctuation power. It has long been known that solar wind fluctuations typically have negative $\sigma_\textrm{r}$ at MHD inertial scales, indicative of a more complex spectrum of turbulence and also non-turbulent magnetic structures rather than one composed purely of Alf\'ven waves. At larger scales, the flux ropes found within interplanetary coronal mass ejections are an example of magnetic structures with negative $\sigma_\textrm{r}$ \citep{Telloni12}. In this work we have focussed on the residual energy of another heliospheric transient, namely the shock wave. Treating shocks as ideal MHD discontinuities conforming to the Rankine-Hugoniot jump conditions, it has been found that the residual energy of fast-mode shocks is positive. In contrast, slow-mode shocks may have positive or negative $\sigma_\textrm{r}$ depending on the Alfv\'en Mach number, while contact and tangential discontinuities have $\sigma_\textrm{r}=-1$. Although framed in terms of interplanetary shock waves, the expressions derived in Section~\ref{sec:equation} also apply, for example, to planetary bow shocks and non-relativistic astrophysical shocks. Shocks with high Mach numbers ($M_\textrm{A}>10$) are expected to have $\sigma_{\textrm{r}}\simeq 1$. Our work also highlights that shock waves can hold a broad range of cross helicity values. In solar wind intervals where they hold a significant fraction of the total fluctuation energy, compressions will complicate the common interpretation of $\sigma_\textrm{c}$ as the imbalance in energy of counterpropagating Alfv\'en waves \citep[e.g.][]{Sishtla24}.

The equation derived for the residual energy of perpendicular shocks broadly agrees with observations. A more comprehensive comparison between theory and observations encompassing parallel and quasi-parallel shocks could be undertaken in future. The near-Sun region now being sampled by the \textit{Solar Orbiter} and \textit{Parker Solar Probe} spacecraft is threaded by a more radial Parker spiral and is thus a more opportune environment for observing quasi-parallel shocks \citep{Good20,Kruparova25,Trotta25}, which are less well represented in the IP Shocks Database currently. Observations of parallel shocks could be used to investigate trends in cross helicity; for example, parallel and quasi-parallel fast-mode shocks with low $M_\textrm{A}$ are expected to have relatively high $|\sigma_{\textrm{c}}|$. Inherent uncertainties in measuring shock parameters may in general make fine distinctions (e.g. relating to the $\theta_{Bn}$ dependence) challenging to test with observations, however. Positive residual energy signatures could play some role in future schemes for identifying fast-mode shocks. 

Together with wavelet spectrograms of the normalized reduced magnetic helicity, $\sigma_{\textrm{m}}$, which give a measure of magnetic field rotations, spectrograms of $\sigma_{\textrm{r}}$ and $\sigma_{\textrm{c}}$ have been used to identify small-scale flux ropes in solar wind data \citep[e.g.][]{Zhao21a,Ruohotie22}. Flux ropes have $\sigma_{\textrm{r}}\simeq-1$ and $\sigma_{\textrm{c}}\simeq0$; given that tangential discontinuities also have these $\sigma_{\textrm{r}}$ and $\sigma_{\textrm{c}}$ values (Section~\ref{sec:other_disc}), consideration of how the magnetic field changes direction (sharply across a tangential discontinuity, more gradually across a flux rope) could help to distinguish between the two types of structure in some ambiguous cases. Furthermore, the $\sigma_{\textrm{r}}\simeq-1$ and $\sigma_{\textrm{c}}\simeq0$ signatures of flux ropes and the tangential discontinuities ideally bounding them \citep[e.g.][]{Ala-Lahti23} would appear as merged features in wavelet spectrograms, and flux rope durations determined from such merged, temporally extended signatures may be overestimates given the temporal broadening at low frequencies associated with large-amplitude discontinuities. 

This work suggests that the presence of fast-mode shocks will increase the overall $\sigma_\textrm{r}$ in an interval, slow-mode shocks may increase or decrease $\sigma_\textrm{r}$, and contact and tangential discontinuities will decrease $\sigma_\textrm{r}$. More generally, compressions are known to affect measurements of $\sigma_\textrm{r}$ and $r_\textrm{A}$ in the solar wind \citep[e.g.][]{Podesta07}. At relatively large scales, \citet{Roberts20} found compressive interaction regions at 1~au to be associated with the highest values of $\sigma_\textrm{r}$ measured across a range of different solar wind types, with $\sigma_\textrm{r}$ in the compression regions often being positive. The positive $\sigma_\textrm{r}$ of these fast-mode-like compressions is in qualitative agreement with our analysis. In the inertial range, \citet{Matthaeus82} found Alfv\'en ratios were usually larger (i.e. closer to unity) when normalizing the magnetic field with the instantaneous, fluctuating density measurements rather than the mean value across an interval. Assuming most solar wind density fluctuations can be characterized as slow-mode-like discontinuities, they could make the total residual energy less negative at $0.5\lesssim M_\textrm{A}\lesssim1$; however, given that these fluctuations typically resemble tangential discontinuities with $M_\textrm{A}\simeq0$, their predominant effect will be to make $\sigma_\textrm{r}$ more negative. Moreover, interpreting the total residual energy measured at inertial-range scales as a simple sum of contributions from different types of MHD discontinuity may not capture the true complexity of the compressive turbulence and its interaction with the Alfv\'enic turbulence. Exploring more broadly the relationship between compressive turbulence and residual energy could be a fruitful avenue of future research; supersonic turbulence close to the Sun, which contrasts with the subsonic turbulence found at larger heliocentric distances and which is expected to produce compressive, shock-like discontinuities \citep[e.g.][]{Zhao25a}, could be of particular interest to examine in light of our analysis.

\section{Conclusion}
\label{sec:conclusion}

The Rankine-Hugoniot conditions have been used to derive equations for the residual energy and cross helicity of an idealized MHD shock in terms of the shock angle, density compression ratio and upstream Alfv\'en Mach number. Simplified forms of these equations valid for the exactly parallel and perpendicular shock have also been derived. Only positive values of residual energy are obtained from these equations for super-Alfv\'enic (i.e. fast-mode) shocks. The values of residual energy predicted by the equation for perpendicular shocks are generally consistent with observations, particularly for shocks with larger $r$ and $M_\textrm{A}$. Fast-mode shock waves have a characteristically positive residual energy signature in observations that stands out markedly from the predominantly negative residual energy spectrum of solar wind fluctuations at MHD inertial scales.

\section*{Acknowledgements}

This work was funded by a Research Council of Finland Fellowship (grants 338486, 346612 and 359914; \mbox{INERTUM}). Recent updates to the IP Shocks Database were supported by a SCOSTEP/PRESTO grant for database development (InnerShock). CHKC and CPS are supported by UKRI Future Leaders Fellowship MR/W007657/1; CHKC is also supported by STFC Consolidated Grant ST/X000974/1. EKJK acknowledges support from the Finnish Centre of Excellence in Research of Sustainable Space (Research Council of Finland grant 352850). Open access to this work has been funded by Helsinki University Library.

SWG wrote the manuscript, derived the results in Section~\ref{sec:equation} and produced Fig.~\ref{fig:fixed_theta}--\ref{fig:slow_shocks}. KJP contributed to the derivation in Section~\ref{sec:equation}, performed the analysis described in Section~\ref{sec:observations} and produced Fig.~\ref{fig:verification_plot}--\ref{fig:wavelet_search}. TVM updated the IP Shocks Database with the shocks analysed in Fig.~\ref{fig:wavelet_plot} and \ref{fig:wavelet_search}. All other authors contributed to the interpretation of results and final preparation of the manuscript. The Python wavelet code used for the analysis presented in Fig.~\ref{fig:wavelet_plot} and \ref{fig:wavelet_search} is found at https://github.com/chris-torrence/wavelets. The perceptually uniform colour map (`batlow') used in Fig.~\ref{fig:fixed_theta}--\ref{fig:verification_plot} was developed by \citet{Crameri21}.

\section*{Data Availability}

Spacecraft data were obtained from the publicly accessible CDAWeb archive at https://cdaweb.sci.gsfc.nasa.gov. The shock database used in this study is found at https://ipshocks.helsinki.fi.

\bibliographystyle{mnras}
\bibliography{references}

\bsp	
\label{lastpage}
\end{document}